# Control and modulation of droplet vaporization rates via competing ferro- and electro-hydrodynamics

**Purbarun Dhar[1], \*, Vivek Jaiswal[2], Hanumant Chate[2]** and **Lakshmi Sirisha Maganti[3]**

[1]Department of Mechanical Engineering, Indian Institute of Technology Kharagpur, Kharagpur–721302, India

[2]Department of Mechanical Engineering, Indian Institute of Technology Ropar, Rupnagar–140001, India

[3]Department of Mechanical Engineering, SRM University–Andhra Pradesh, Amaravati–522503, India

\* *Corresponding author*:

E–mail: purbarun.iit@gmail.com ; purbarun@mech.iitkgp.ac.in

Phone: +91–1881–24–2119

## Abstract

Modification and control over the vaporization kinetics of microfluidic droplets may have strong utilitarian implications in several scientific and technological applications. The article reports the control over the vaporization kinetics of pendent droplets under the influence of competing internal electrohydrodynamic and ferrohydrodynamic advection. Experimental and theoretical studies are performed and the morphing of vaporization kinetics of electrically conducting and paramagnetic fluid droplets using orthogonal electric and magnetic stimuli is established. Analysis of the observations reveals that the electric field has a domineering influence compared to the magnetic field. While the magnetic field is noted to aid the vaporization rates, the electric field is observed to decelerate the same. Neither the vapour diffusion dominated kinetics nor the field induced modified surface tension can explain the observed vaporization behaviours.

Velocimetry within the droplet shows largely modified internal ferro and electrohydrodynamic advection, which is noted to be the crux of the mechanism towards modified vaporization rates. A mathematical treatment is proposed and takes into account the roles played by the governing Hartmann, electrohydrodynamic, interaction, the thermal and solutal Marangoni, and the electro and magneto Prandtl and Schmidt numbers. It is observed that the morphing of the thermal and solutal Marangoni numbers by the electromagnetic interaction number plays the dominant role towards morphing the advection dynamics. The model is able to predict the internal advection velocities accurately. The findings may hold significant promise towards smart control and tuning of vaporization kinetics in microhydrodynamics transport paradigms.

# 1. Introduction

Hydrodynamics and thermo-species transport phenomena in microscale droplets has been an area of interest to researchers and academicians owing to the exotic physics involved and the plethora of technological utilities associated with the phenomena. A wide gamut of applications of droplets and sprays exist, such as in biomedical devices like nebulizer and inhalers [1], fuel injection system in combustion engines [2-3], thermal treatment and printing technology [4-5], and in the HVRAC sector. Fumigation by insecticides and repellents [6-7] and microfluidic organ-on-chip devices [8-9], and biological sorting devices [10] are other important applications of droplet hydrodynamics. Typically, research on droplets is inclined towards sessile droplets. Fluid dynamics and thermo-species transport in such droplets has been extensively looked into. The more prominent works, such as reports by Hu and Larsen [4], Semenov et al. [11], Bekki et al. [12] and Fukai et al. [13], etc. are noteworthy. Nevertheless, in recent times, the research on microscale droplets has seen a paradigm shift to pendant droplets. This has been triggered by the fact that pendant droplets are independent of the surface interactions that sessile droplets experience, thereby allowing to mimic the transport phenomena in free-standing droplets.

The bulb of the pendent droplet assumes a near-spherical shape and the thermo-hydrodynamic processes can be translated to that within a spherical free-standing droplet. Godsave [14] analysed the vaporization of such a droplet and introduced a linear relationship between the square of the non-dimensional diameter and the elapsed modified droplet lifetime. This proposed classical $D^2$ law was further confirmed experimentally for several vaporization conditions and ambient characteristics [15]. In the recent era of multiphysics, hydrodynamics and transport phenomena have been shown to be tuned, modified and subject to smart utilities by the use of non-tactile force fields. Among them, the electrohydrodynamic and ferrohydrodynamic routes have gained popularity due to the relative ease of implementation and the flexibility of tuning the thermo-hydrodynamic behaviour. Some recent developments in this direction discuss the role of magnetic field on the thermo-hydrodynamics in magnetic fluids [16, 17]. The presence of external magnetic field can create internal ferro-advection within paramagnetic fluid droplets, which can enhance their vaporization rates [18]. Unlike normal fluids, magnetic fluids experience the Kelvin force and the body couple [19] in the presence of a magnetic field. Rossow [17] reported modulated thermal transport of an electrically conducting solution droplet in the presence of a magnetic field. Sparrow and Cess [20] showed natural convection behaviour of magnetic fluids is drastically modified due to the magnetic body force.

Nusselt numbers during free convection of electrically conducting fluids confined within enclosures was reported to decrease due to morphed flow dynamics at various Hartmann numbers [21]. Along similar lines, external magnetic field was shown to change hydrodynamics and transport behaviour in parallel shear flows [22]. Magyari et al. [23-24] and Kandasamy et al. [25] compiled comprehensive details on analytical and simulation reports and concluded that magnetic fields can be used effectively to modulate flow dynamics and heat transport in electrically active fluids. The presence of magnetic field has been shown to lead to thermo-capillary Marangoni convection [26, 27] in fluidic interfaces. Ferrofluids have also been shown to undergo changed dynamics during Rayleigh-Benard convection in the presence of magnetic field [28]. Singh and Bajaj [29] and Belyaev and Smorodin [30] investigated ferro-convection behaviour with sinusoidal thermal stimulus, and alternating magnetic field, respectively, and simulation results show that the ferrohydrodynamics is strongly dependent on the field and the thermal constraints. Likewise, reports show that Marangoni ferro-convection in ferrofluids is strongly dependent on the thermal gradients within the system as well as the strength of the local magnetic field [31, 32]. Jaiswal et al. [33] showed experimentally and theoretically that the fluid dynamics and thermal transport in paramagnetic droplets can be modulated by magnetic field, which leads to augmented evaporation of such droplets.

Electric field has also been shown to modulate hydrodynamics and associated thermal transport behaviour. The shape of the droplet is governed by surface tension, gravity, and (in presence) by the electric field also. The interfacial characteristics also vary due to the application of electric field, leading to morphed interfacial hydrodynamics. The droplet shapes under electrostatic forces have been studied extensively, the more important reports being by Bateni et al. [34-35], Reznik et al. [36], Basaran and Wohlhuter [37] and Miksis [38]. The electrical forces within hydrodynamic systems may be combination of the Coulomb, electrokinetic, dielectrophoretic, or electrostriction forces. In the presence of electric field stimulus, the internal and interfacial hydrodynamics of conducting solution droplets are and modified, which leads to suppressed mass transfer rates [39]. In the presence of strong field intensities, the droplets exhibit interfacial deformation, which have been modelled by incorporating the Maxwell stresses to the interfacial hydrodynamic stress [40]. Different electrohydrodynamic paradigms, such as electrospinning, electro-Leidenfrost impact, electro-propulsion, and electro-wetting have been studied widely in recent times [41-44]. Modulation of the slip-stick behaviour of sessile droplets by interplay of wettability and electric field has also been shown to lead to self-propelling sessile droplets [45]. Electric fields can prompt the formation of Taylor cones at microfluidic junctions which can be utilized to create homogenized emulsions [46]. Electro-coalescence, electro-separation and electro-deposition of droplets [47, 48], and microfluidic electro-sorting of malignant biological cells [49] are other typical systems which have been studied under the regimes of droplet electrohydrodynamics [50].

The present article aims to probe into the physics and mechanics of vaporization of pendent droplets in the presence of interplaying electrohydrodynamics and ferrohydrodynamics. Previous works by the authors have shown that paramagnetic saline solution droplets exhibit enhanced internal advection, and thereby improved vaporization in the presence of magnetic field [33]. The authors also showed that in conducting saline droplets, the internal advection and the vaporization rates are reduced due to the presence of electric field [39]. Thus, the two components of the electromagnetic Lorentz force are noted to have distinctly different effects on the hydrodynamics and the thermo-species transport in droplets. However, the Lorentz force contains within itself an interaction of the orthogonal electric and magnetic fields, whose effects on the hydrodynamics is expected to reveal intricate physics, and may bear far reaching implications in microfluidic manipulation and applications. This study focuses on the vaporization kinetics of a paramagnetic saline solution droplet and the associated interacting internal electro and ferrohydrodynamics in the presence of orthogonal electric and magnetic fields. The study provides concrete experimental findings and a theoretical scaling analysis to establish the sensitivity of individual fields on the thermo-species transport and the role of the interaction component of the Lorentz force on the associated hydrodynamics.

## 2. Experimental methods

In the present study, anhydrous iron (III) chloride ($FeCl_3$, Sigma Aldrich, India) based paramagnetic aqueous solution has been used. The advantage of the salt is that it provides both a paramagnetic and a conducting medium at the same time, which responds to both magnetic and electric fields. Two different concentrations of 0.1 M and 0.2 M are used. Fig. 1 illustrates the experimental setup employed. The droplet was generated at the tip of a stainless steel needle using a digitally controlled precision dispensing mechanism (Holmarc Opto-mechatronics Pvt. Ltd., India). A glass syringe of capacity $50 \pm 0.1$ μL has been used, and a droplet of $10 \pm 0.5$ μL was suspended as a pendant. The diameter of the droplet was measured to be in the range of 2.5 $\pm$ 0.2 mm. The electric field was generated by a programmable AC power supply (50 Hz) (Aplab, India) with field range of 0-270 V and 0.1 V resolution. One terminal was connected to the needle suspending the droplet. The other terminal was connected to a steel sheet placed horizontally below the hanging droplet (~1 mm away from the droplet base). The absolute separation between needle electrode and the steel plate electrode was maintained as 3.5 ± 0.1 mm.

The magnetic field was generated across the droplet by a horizontal pole electromagnet with digitized current control (Polytronic Corporation, India). The droplet was suspended vertically from the needle at the centre of the electromagnetic pole shoes. The pole separation was maintained at 20 mm, and the electromagnet was initially calibrated in this configuration using a digitized Gaussmeter (with an InAs based sensor). The electromagnet has a field range of 0-1 T. Thermal imaging shows that at 0.3 T Ga and beyond, the electromagnet poles start to show mild Joule heating. This behaviour can influence the evaporation rates and hence the magnetic field strength has been limited to 0.2 T. A charged coupled diode (CCD) camera (Holmarc Opto-mechatronics Ltd., India) with a long distance microscopic lens was used to record the vaporization phenomena. The camera was operated at 1280 x 960 pixels resolution and frame rate of 10 Hz. A brightness controlled LED was used as the illumination source. The complete setup was housed inside an acrylic chamber to eliminate ambient and human disturbances. A digital thermometer and a digital hygrometer were used to measure the ambient temperature and humidity conditions, respectively. The sensing probes of these devices were positioned ~ 20 mm away from the droplet.

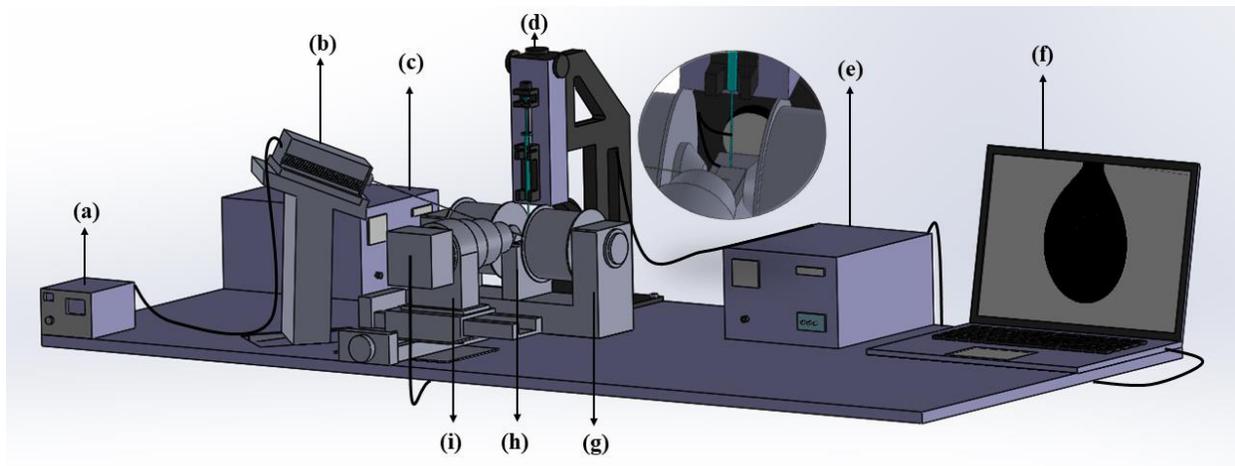

**FIG. 1:** Schematic of the experimental setup and components (a) laser controller (b) laser mounted on stand (for PIV) with cylindrical lens assembly (not shows) (c) electromagnet power supply and controller (d) droplet dispensing mechanism and backlight illumination attachment (e) programmable AC power supply unit (f) data acquisition and camera control computer (g) electromagnet unit (h) mounting unit for the electrode assembly (i) CCD camera with long distance microscope lens assembly and two-axis movement control. Components (b), (d), (g), (h) and (i) are enclosed in an acrylic chamber. The inset shows a zoomed view of the electrode connections (vertical electric field) and the placement of the needle between the magnetic poles (horizontal magnetic field).

The temperature and humidity were noted as $28 \pm 2$ $^o$C and $55 \pm 4$ % for the complete set of experiments. The images obtained were post-processed in the open-source software ImageJ and the instantaneous diameters of the droplets were determined. The evaporation rate of deionized water serves as the benchmark, and at zero electric fields and zero magnetic fields, the values agree with literature reports [18, 33, 39]. Particle Image Velocimetry (PIV) was performed to qualitatively and quantitatively describe the internal hydrodynamics of the droplet. Neutrally buoyant fluorescent particles (polystyrene, ~10 µm, Cospheric LLC, USA) were used as the seed particles. The excitation source for the PIV is a continuous-wave laser (Roithner GmbH, Germany, wavelength 532nm and 10 mW peak power). A laser sheet (~0.5 mm thickness) is generated using a cylindrical lens and the mid-plane of the droplet is optically probed. 120 pixels/mm resolution was used for the velocimetry performed at 30 fps. The data was processed for typically 800 images for each case using the open-source code PIVlab. A 4 pass cross-correlation algorithm (interrogation windows of 64, 32, 16 and 8 pixels) was used to process the PIV data to obtain high signal-to-noise ratio. The average displacement of the particles is a function of the salt concentration, nature and strength of fields applied. However, for the 0.2 M solution droplet under zero-field, the average displacement of particles between frames was noted as ~ 0.25 mm.

## 3. Results and discussions

### 3. a. Evaporation in the presence of conjugate electromagnetic stimulus

Figure 2 (a) and (b) illustrates the role of the magnetic field on the evaporation characteristics of the droplets when the electric field is kept constant. The rate of evaporation ($k$) is obtained from the D$^2$ law, which is expressed as [33, 39]

$$\frac{D^2}{D^2_0} = 1 - k\frac{t}{D_0^{\ 2}} \tag{1}$$

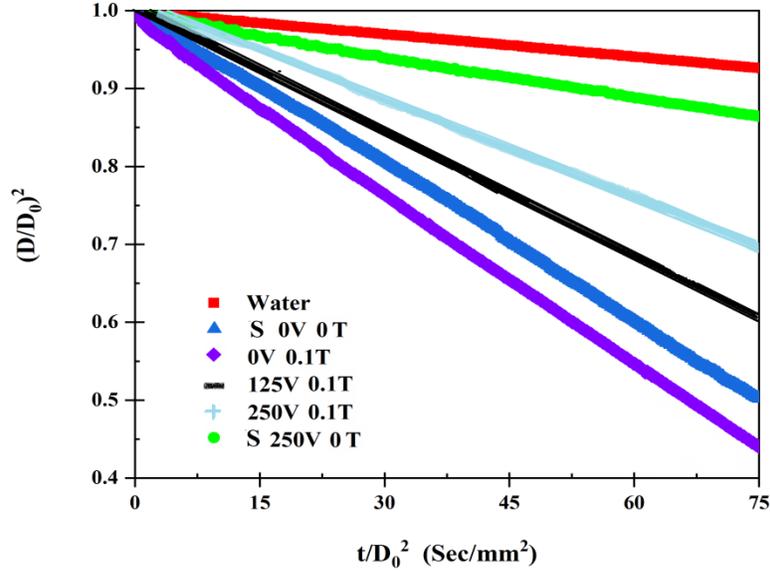

**FIG. 2:** Evaporation characteristics of water and saline droplet of 0.1 M concentration (represented by S) at electric field of 125V and 250 V, for magnetic field strength of 0.1 T. The water and S are at 0 V and 0 T in both cases.

In eqn. 1, D is the instantaneous droplet diameter, $D_0$ is the initial diameter of the droplet, and $t$ is the elapsed time. Coming back to fig. 2, it is known from reports that addition of salt enhances the evaporation rate of water droplets due to internal thermo-solutal advection and morphed Stefan flow due to the interfacial shear [18]. Now, as the magnetic field increases, the evaporation rate improves, and this is already reported by the authors [33]. In the case of only electric field however, increase of field strength leads to reduction in the evaporation rates [39]. In fig. 2 (a) and (b), the saline droplet exhibits largely improved evaporation rate. With the application of electric field, the evaporation rate reduces. Further, with the application of conjugate magnetic field, the evaporation rate improves, but even with 0.1 T, cannot recover back to the case of saline droplet. But in reality, only the magnetic field would lead to further increase in evaporation rate of the saline droplet [33], as shown in the fig. 2 (a). Hence it is noted that in the case of conjugate orthogonal fields, the electric field has a typically dominance over the magnetic field. Figure 3 illustrates the evaporation behaviour of 0.2 M $FeCl_3$ solution droplet under different field configurations and the objective is to show the effect of magnetic field at constant electric field. In the absence of magnetic field, the electric field leads to reduction in the evaporation rate of the saline droplet, which is consistent with reports [39]. The magnetic field in itself leads to enhancement of the evaporation rate of the saline droplet [33]. However, the conjugate field evaporation rates are always smaller than that of the saline droplet, even in the case of weak electric field and strong magnetic field (125 V, 0.2 T). This furthers the observation that in case of orthogonal fields, the electric field shows a dominant effect over the magnetic field. Hence, it can be inferred from figs. 2 and 3 that a wide range of droplet vaporization rates can be achieved and realized by tuning the salt concentration, electric and magnetic field strengths. This can have tremendous potential in microfluidic systems.

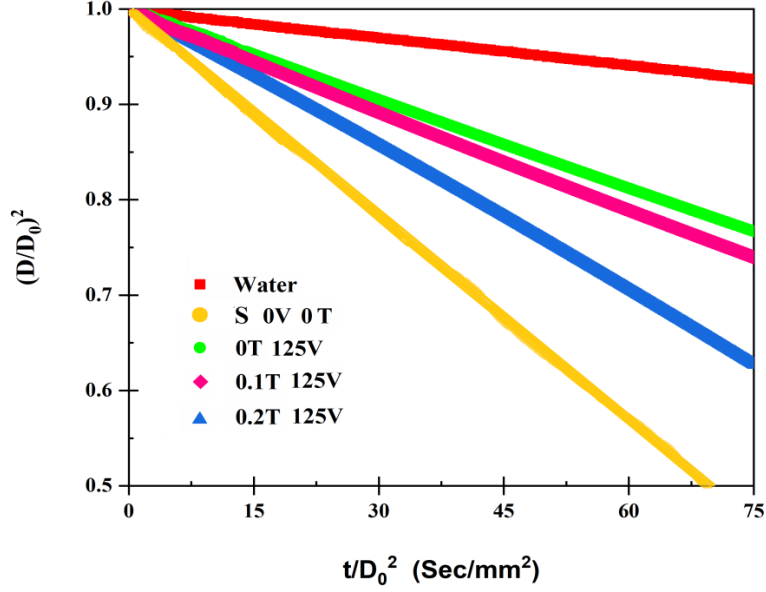

**FIG. 3:** Evaporation characteristics of water and saline droplet of 0.2 M concentration (represented by S) at magnetic field of 0.1 T and 0.2 T, for electric field strength of 125 V. The water and S are at 0 V and 0 T in both cases.

### 3. b. Role of the vapour diffusion layer

Pendant droplets evaporate due to diffusion dominated mode, wherein the fluid molecules escape the liquid surface to the vapour diffusion layer shrouding the droplet. The vapour molecules then diffuse from the diffusion layer to the ambient by virtue of the concentration gradient. The diffusion-driven evaporation kinetics can be mathematically modelled to predict the evaporation rates [51]. The classical model considers the concentration difference across the diffusion layer around the droplet and the ambient air, and deduces the evaporation rate from equations 2 to 5 as [51]

$$B_M = \frac{Y_s - Y_\infty}{1 - Y_s} \tag{2}$$

$$B_T = (1 + B_M)^\Phi - 1 \tag{3}$$

$$\phi = \frac{C_{pf} Sh}{C_{pg} NuLe} \tag{4}$$

$$\overset{\bullet}{m} = 2\pi \rho_g D_v R \ln(1 + B_M) Sh = \frac{2\pi \lambda R}{C_{pf}} \ln(1 + B_T) Nu \tag{5}$$

In eqns. 2-5, $B_M$, $Y_s$ and $Y_\infty$ are the Spalding mass transfer number, mass fraction of the vapor at the droplet surface, and in the ambient phase. $B_T$ is the Spalding heat transfer number and $C_{pg}$, $Nu$, $Le$, and $Sh$, are the specific heat of the vapour film, the specific heat of the surrounding gas, the Nusselt number, Lewis number and Sherwood number, respectively.

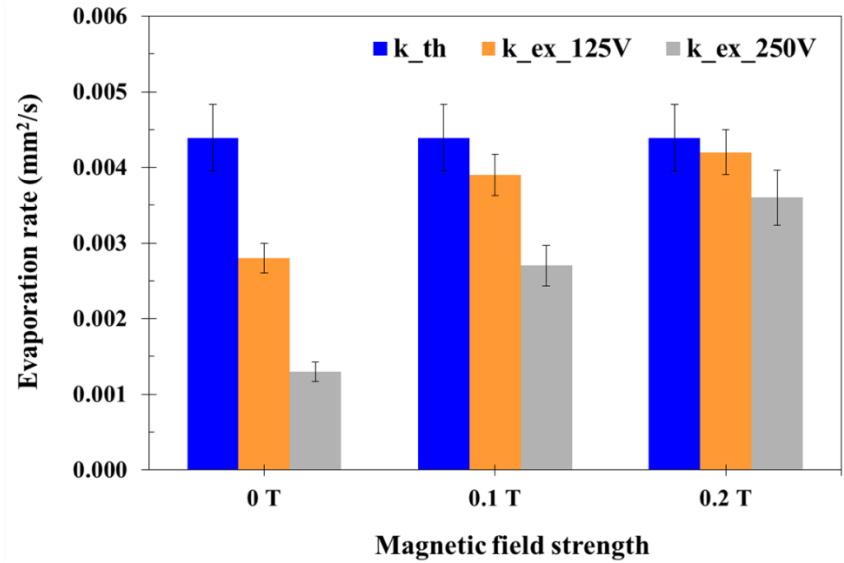

**FIG. 4:** Comparison of experimental evaporation rates with the diffusion-driven theoretical approach for different electric and magnetic field strengths for 0.2 M saline droplets.

In eqn. 6, $\dot{m}$ is the mass loss rate by evaporation, $D_v$ is the diffusion coefficient of the vapour with respect to the ambient phase, $\rho_g$ is the density of the ambient gas, R is the droplet instantaneous radius, and $\lambda$ is thermal conductivity of the ambient gas phase. Fig. 4 illustrates the comparison of experimental evaporation rates with the diffusion driven model predictions, for different electric and magnetic field strength for 0.2 M FeCl$_3$ solution droplets. It is noted that the model cannot predict the evaporation rates, and yields a constant value. This is however not surprising, as the model considers the gas phase dynamics only, and in the present case the external conditions remain unaffected. Thereby, the model is unable to predict the changes in the evaporation dynamics, and the exercise shows that the crux of the mechanism is not on the exterior, but either at the droplet surface, or in its interiors.

### 3. c. Surface tension modulation under field stimulus

One mechanism by which the evaporation kinetics of droplets may be morphed is by the change in surface tension of the fluid. Typically, fluids with low surface tension, such as alcohols and volatile oils, are known to vaporize faster due to the ease of escaping the liquid-gas interface. Figure 5 illustrates the surface tension of 0.2 M FeCl$_3$ solution under the effect of different magnetic and electric field strengths. The surface tension is determined using the pendent drop method [18] and the pure deionized water exhibits a surface tension of 69 ± 2 mN/m. It is noted from the figure that at low electric field strengths, the surface tension improves with increasing magnetic field. However, this increase is restricted in magnitude for the case of high electric field strengths. However, no generic trend is noted and the modulation in surface tension is not proportional to the observed changes in the evaporation rates, which is also consistent from literature [33, 39]. Consequently, surface tension modulation is not a potent method to explain the morphed evaporation kinetics, and probing within the droplet is deemed essential.

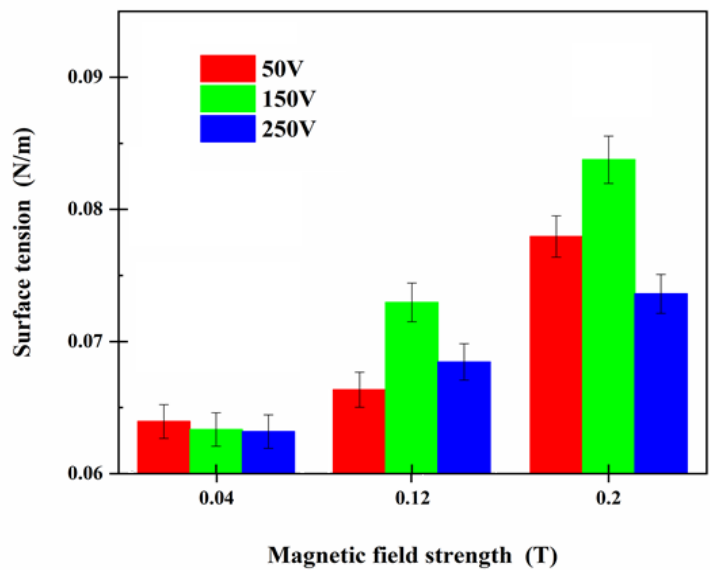

**FIG. 5:** Surface tension of the saline solutions (0.2 M concentrations) for different magnetic and electric field strength.

### 3. d. Internal hydrodynamics due to field stimulus

PIV studies have been performed to observe and quantify the internal advection dynamics. The PIV were conducted during the initial five minutes of the evaporation of respective experiments, such that the effect of change of salt concentration on the internal hydrodynamics can be restricted to a minimum. The genesis and kinetics for the internal hydrodynamics within such saline droplets has been discussed in previous reports by the authors [18, 33, 39]. The velocity of the circulation within water and saline droplets (zero field condition) is in agreement with with reports in the literature [18, 33]. For the case of water (both with and without field), insignificant internal circulation was observed, and very minor drift was noticeable (average velocity <0.1 cm/s). Fig. 6 illustrates the velocity contours and vector fields (temporally averaged over 90 seconds) within 0.2 M droplets for different electric and magnetic field strengths. The neck of the droplet is removed from consideration to avoid the possible spurious flow effects due to the presence of the needle. In fig. 6 (a), the case of 0.2T and 0 V has been illustrated. The velocity vectors within the droplet mid-plane are in the same direction as the magnetic field lines (horizontal). The magnitude of the advection is also higher compared to the 0 T 0 V case, which is in agreement with literature [33]. Likewise, fig. 6 (d) illustrates the case of 250 V and 0 T, and the advection is noted to be largely suppressed compared to the 0 V 0 T saline case, with minimal fluid motion confined towards the neck of the droplet. The weak advection is also directed from the bottom towards the top of the droplet. This is also in agreement with literature [39].

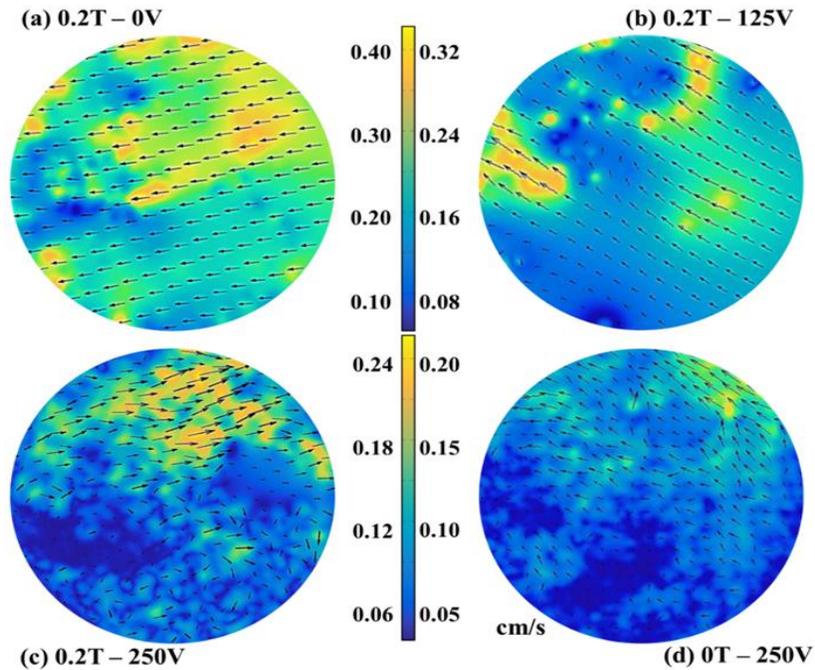

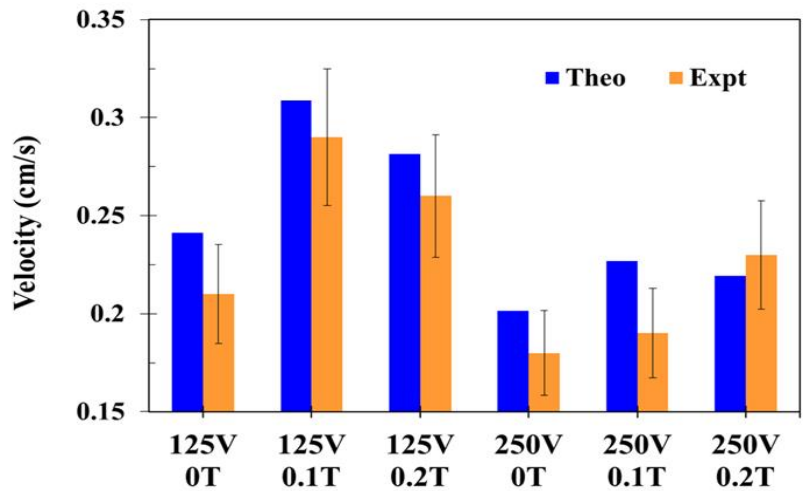

**FIG. 6:** Internal advection velocity contours and time-averaged flow field for 0.2 M droplet for (a) 0.2 T–0 V (b) 0.2T–125V (c) 0.2T–250V and (d) 0T–250V (e) comparison of predicted velocities from the proposed scaling model with experimental velocities

Fig. 6 (b) and (c) illustrates the cases with orthogonal electric and magnetic fields acting simultaneously across the droplet. In 6 (b), an electric field of 125 V is added over the magnetic field in 6 (a). In 6 (a), the advection was directed along the magnetic field lines, however, with the introduction of the electric field, the direction of advection changes partly towards the vertical electric field lines. In addition, the average strength of advection also reduces. In 6 (c), the electric field strength is further stepped up to 250 V at the same magnetic field strength. It is noticed that the advection pattern is now confined largely towards the top of the droplet (very similar to 6 (d)), albeit with higher advection strength compared to 6 (d). The PIV studies clearly show that the competing electrohydrodynamics and ferrohydrodynamics lead to the final advection patterns within the droplet. Studies [18, 33, 39, 50] have established that this advection induces shear on the droplet–vapour interface, which in turn shears the vapour diffusion layer shrouding the droplet. This shear replenishes the otherwise stagnant vapour diffusion layer with the ambient air, thereby enhancing the Spalding mass transfer number, and improving the evaporation rates. In the present case, the different advection strengths lie in between the cases of only magnetic field and only electric field, and accordingly the vaporization rates can be tuned to any value within these two limits. However, the genesis of this combined advection, whether driven by the electromagnetic–thermal effects or the electromagnetic–solutal effects remains to be probed.

### 3. f. Role of electro-magneto-thermal advection

One mechanism by which the internal advection is generated within the vaporizing droplet is by the thermal Marangoni effect, wherein the thermal gradients developed in the droplet during evaporation [52] drives the surface tension gradient driven hydrodynamics. In the presence of the electromagnetic field, this hydrodynamics is morphed into what may be described as an electro-magneto-thermal advection. Considering energy balance across the vaporizing droplet, and the associated modes of thermal transport within the droplet, a conservation equation may be established [18] as

$$\dot{m} h_{fg} = K_{th} A \frac{\Delta T_m}{R} + \rho C_p U_{c,m} A \Delta T_m + \rho C_p V_f A \Delta T_m \tag{6}$$

Where, $\dot{m}$, $h_{fg}$, $K_{th}$, $C_p$, $\rho$ and $A$ refers to change in mass rate of vaporization, the enthalpy of vaporization, the thermal conductivity of the liquid, the specific heat of the liquid, density of the liquid and heat transfer area of the droplet, respectively. $\Delta T_m$, $V_f$ and $U_{c,m}$ represent the temperature difference across the droplet bulk and interface, created due to vaporization, the spatially averaged internal electro-magneto-thermal advection velocity, and the spatially averaged internal velocity due to thermal advection, respectively. The left side of the energy equation represents the energy flux due to the vaporization mass loss, and the right side includes sum of the heat conduction across the droplet, the convective heat flow within the droplet driven by temperature gradient, and the convective flow due to the electro-magneto-thermal stimulus, respectively.

The average internal advection velocity generated by the thermal Marangoni effect is expressible as, $U_{c,m} = \sigma_T \Delta T_m / \mu$ [52], where $\sigma_T$ represents the rate of change of surface tension with change of temperature and $\mu$ is the dynamic viscosity of the liquid. The presence of the solvated paramagnetic ions in the fluid leads to Lorentz force within the droplet. For a moving charge inside a coupled electromagnetic field, the net force experienced by the charged body is expressible as,

$$\vec{F} = \rho_e \vec{E} + (\sigma_e . \vec{E} \times \vec{B}) + \sigma_e (\vec{v} \times \vec{B}) \times \vec{B} \tag{7}$$

The magnitude of this force may be scaled in a simplistic mannerism as

$$|F| = \rho_e E + \sigma_e E B + \sigma_e U_{c,m} B^2 \tag{8}$$

The electromagnetic force gives rise to inertia within the liquid, which in turn accelerates or decelerates the thermal advection. The inertia force can be expressed in terms of the acceleration ($a$) experienced by the fluid element, leading to the expression,

$$F = \rho a \tag{9}$$

The acceleration can be further scaled as $a \sim V_f / t$ (where $t$ is a characteristic time period, which can be further scaled as $t \sim R/V_f$), and thus the acceleration scales as $a \sim \rho V_f^2 / R$. Equating the forces for equilibrium, one obtains

$$V_f^2 = E_{hd} \frac{v^2}{R^2} + \frac{\sigma_e E B R}{\rho} + Ha^2 U_{c,m}^2 \frac{v^2}{R^2} \tag{10}$$

where, $v$ and $E_{hd} = eNZER^3/\rho v^2$ represent the kinematic viscosity and the Electro-hydrodynamic number, respectively. The non-dimensional $E_{hd}$ represents the ratio of the electric force to the viscous forces in an electrohydrodynamic system. The non-dimensional Hartmann number $\left( Ha = BR\sqrt{\sigma_e/\mu_f} \right)$ is the ratio of the electromagnetic Lorentz force to the viscous force in a magnetohydrodynamic system.

Further simplification of the eqn. 10 yields

$$V_f^2 = E_{hd} \left( \frac{v}{R} \right)^2 + Ha.I \left( \frac{v}{R} \right) U_{c,m} + Ha^2 \left( \frac{v}{R} \right)^2 U_{c,m}^2 \tag{11}$$

where $I = ER/U_{c,m}\sqrt{\sigma_e/\mu_f}$ is defined as the Interaction number, and represents the interaction between the electric and magnetic fields within the hydrodynamic system. Now, the velocity can be scaled as $V_f \sim U_{c,m}$, and further observation shows that for a droplet system such as the present one, $Ha \sim 1$, and $(\nu/R) \ll 1$. Employing these scaling to eqn. 11, the expression for the electro-magneto-thermal advection velocity can be obtained as

$$V_f = \frac{\nu}{2R}\left\{ HaI \pm \left(\sqrt{Ha^2 I^2 + 4E_{hd}}\right)\right\} \tag{12}$$

In equations 7-12, $E$, $B$, $\sigma_e$, $\mu_f$, $t$ and $R$ represents electric field intensity, magnetic field strength, the electrical conductivity of the fluid, dynamic viscosity of fluid, time, and the radius of the droplet, respectively. Substituting the expressions for $U_{c,m}$ and $V_f$ in eqn. (6) yields

$$\rho \dot{R} R h_{fg} = k_{th}\Delta T_m + \rho C_p \left(\frac{\sigma_T \Delta T_m}{\mu}\right)\Delta T_m R + \rho \frac{C_p \nu}{2}\left\{ HaI \pm \sqrt{(HaI)^2 + 4E_{HD}}\right\}\Delta T_m \tag{13}$$

As stated in literature [18], the thermal gradients appear at the droplet surface, as well as in the bulk phase, leading to both surface and bulk advection. However, since the quantification of surface advection by experiments is challenging, the Marangoni number approach enables to correlate it to the bulk advection (from the very definition of the Marangoni number).

The associated thermal Marangoni number ($Ma_T$), which governs the advection within the droplet solely due to temperature induced Marangoni effect can be expressed as [18]

$$Ma_T = \frac{R}{\alpha}\sqrt{\frac{\dot{R} h_{fg}\sigma_T}{C_p \mu}} \tag{14}$$

Substitution of eqn. 14 in eqn. 13, and rearrangement yields

$$\rho \dot{R} R h_{fg} = k_{th}\Delta T_m \left\{ 1 + Ma_T + \frac{Pr}{2}\left( HaI \pm \sqrt{Ha^2 I^2 + 4E_{hd}}\right)\right\} \tag{15}$$

where $Pr$ represents the Prandtl number. $Ma_T \gg 1$ is an essential criterion for stable internal advection within the droplet [18, 51], and the expression can be simplified to

$$\rho \dot{R} R h_{fg} = k_{th}\Delta T_m \left\{ Ma_T + \frac{Pr}{2}\left( HaI \pm \sqrt{Ha^2 I^2 + 4E_{hd}}\right)\right\} \tag{16}$$

Comparison with the thermal advection model [18] shows that the term $Ma_T + \frac{Pr}{2}\left( HaI \pm \sqrt{Ha^2 I^2 + 4E_{hd}}\right)$ essentially behaves as the effective electro-magneto-thermal Marangoni number ($Ma_{T,m}$), which is similar in concept as the magneto-thermal and electro-thermal Marangoni numbers [33, 39]. Hence, it is the modulation of the interfacial and internal advection behaviour by the fields that is responsible towards the morphed vaporization features.

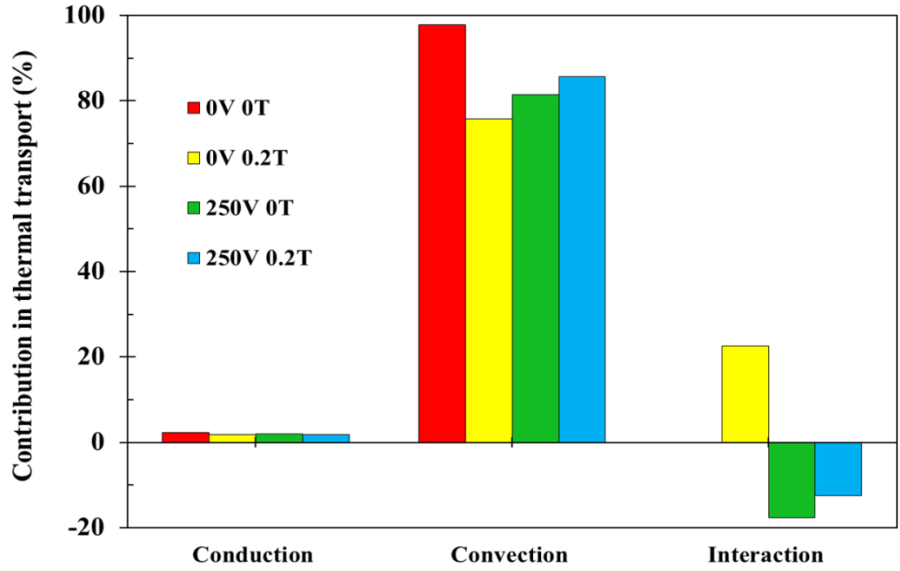

**FIG. 7:** The typical percentage contribution of the different mechanisms involved in the energy transfer during the vaporization of 0.2 M droplet in different electric and magnetic field constraints.

Based on the RHS of eqn. 13, the typical contribution of each mode of thermal transport can be quantitatively estimated. In fig. 7 these contributions to the energy transport during the vaporization have been shown. It is observed from fig. 7 that the conduction mode is negligibly small and the thermal advection component is the strongest mode in the absence of external fields. In the case of only magnetic field (0V 0.2 T), the thermal convection component reduces and the field interaction component (with only the magnetic part active) plays ~ 20 % role towards the net transport process, and is similar to previous report on ferrohydrodynamics in evaporating droplets [33]. In the case of only electric field (250 V 0 T), the field interaction component (with only the electric part active) is ~ 20 % responsible for the transport phenomena (the negative sign portrays the role of decelerated evaporation in case of only electric field [39]). In the case of interacting fields (250 V 0.2 T), the interaction component still remains high, with some reduction in negative magnitude due to the aiding role of the magnetic field. However, overall it typically possesses 15–20 % contribution to the thermal transport phenomena during the vaporization process. Further, the possibility of morphed vaporization rates can also be linked to changes in the nature of the Rayleigh convection within the ambient gas phase. However, analysis [52] has shown that it is very weak compared to the improvement in vaporization caused by the internal circulation within the droplet. Thereby, this mechanism is deemed negligible.

Internal advection due to thermal stimulus could also be due to buoyancy gradients, or Rayleigh advection. The circulation velocity induced within the droplet by buoyancy is expressed [52, 18] as

$$u = \frac{g\beta\Delta T_R R^2}{\nu} \tag{17}$$

and the associated Rayleigh number is expressible as

$$Ra = \frac{R^2}{\alpha}\sqrt{\frac{\dot{R}h_{fg}g\beta}{C_p\nu}} \tag{18}$$

where, the temperature difference inducing this buoyant convection within the droplet can be expressed as

$$\Delta T_R = \sqrt{\frac{\nu\dot{R}h_{fg}}{g\beta R^2 C_p}} \tag{19}$$

In eqns. 17-19, $\Delta T_R, \beta, \alpha$ ,g and $\nu$ represent the temperature difference driving the buoyant advection, coefficient of thermal expansion of the liquid, thermal diffusivity of the liquid, acceleration due to gravity, and kinematic viscosity of the liquid. However, the root of the thermal advection, whether due to Marangoni effect or buoyancy gradients, is to be established. This is determined from the stability of advection criteria in terms of the $Ra_c$ (the critical Ra) and $Ma_c$ (critical thermal Ma). The criterion for stable advection is given as [32, 53]

$$\frac{Ra}{Ra_c} + \frac{Ma}{Ma_c} = 1 \tag{20}$$

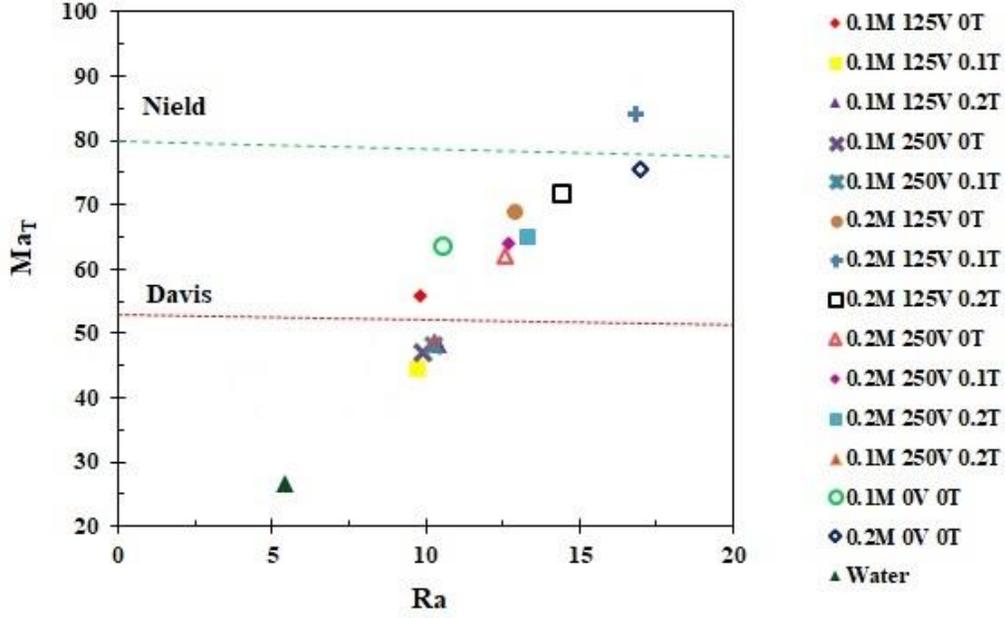

**FIG. 8:** Stability map between the Ra and thermal Ma under different electric and magnetic field stimuli. The lines labelled as Nield and Davis represents the criteria for stable advection [32, 53].

Figure 8 illustrates a stability map for $Ma_T$ and $Ra$ wherein the various vaporizing droplet configurations have been plotted. The two stability criteria by Nield and Davis [32, 53] have been shown as lines. For both conditions, the $Ra_c$ conforms to the analysis by Chandrasekhar (~1708), whereas the $Ma_c$ is ~ 81 for Nield, and ~53 for Davis. The region in between corresponds to intermittently stable advection, whereas the region above the Nield line signifies stable internal advection. It is noted that the water case lies within the regime of unstable circulation, and salt cases (zero field) lie in the region of stable or intermittently stable circulation, which are confirmed from the PIV studies. However, the different field constraint based cases lie within the region of intermittent or stable circulation, although several such cases illustrate reduced advection characteristics in the visualization exercise. Thereby, the thermal Ma vs. Ra is not a proper phase plot to determine the real mechanisms at play. Consequently, the effective electro-magneto-thermal Ma ($Ma_{T,m}$) is plotted against the Ra (in fig. 9) to understand the role of the field modulated thermal Marangoni advection. Figure 9 shows the phase plot for $Ma_{T,m}$ vs. Ra, and the different points have been illustrated in the phase map. It is noted that in this phase plot, the points show the proper trend of shift from the zone of stable advection to that of unstable advection, under the influence of different field constraints. These are in agreement with the PIV observations. Representative iso-$E_{hd}$ and iso-$Ha$ lines have been illustrated in fig. 9 to show the effect of the electric and magnetic fields on the advection regimes. It is noted that with increasing Ha (only magnetic field), the iso-Ha lines shift in a manner where the points shift towards the stable advection regimes and beyond. In the case of the increasing $E_{hd}$ (only electric field), the iso-$E_{hd}$ lines shift in a manner that the points transit towards the region of intermittent to unstable advection. Both these are in agreement with observations and literature reports [33, 39]. Thus the exercise shows that the thermal advection is weakened by the electro-magnetic fields, however, whether the reduction in vaporization kinetics is driven completely by the thermal component, or the solutal component is also responsible, remains to be probed.

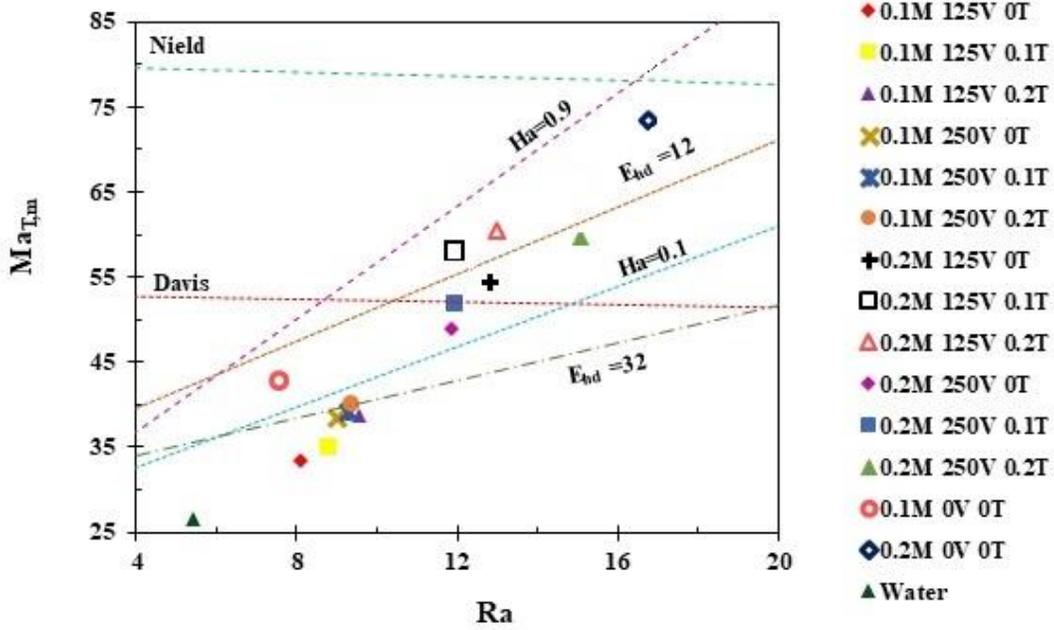

**FIG. 9:** Phase plot of effective electro-magneto-thermal Ma and Ra for different droplets under different electric and magnetic field constraints. The iso-$E_{hd}$ and iso-$Ha$ lines have been illustrated to show the role of the effect of electric and magnetic field strength on the stability regimes [33, 39].

### 3. g. Role of electro-magneto-solutal advection

The internal advection is also generated in saline droplets by solutal Marangoni advection [18, 52], and the morphing of the solutal advection by the electro-magnetic fields may also be a dominant cause for the modulated vaporization features. The genesis of this solutal advection has been discussed in previous reports by the authors [18, 33, 39] and has not been repeated here for the sake of brevity. Similar to the energy balance scaling, the species transport mechanisms due to diffusion, solutal advection and electro-magneto-solutal advection during the vaporization process is scaled, and the conservation equation is expressed as

$$\dot{m} = DA\frac{\Delta C_m}{R} + U_{c,m}A\Delta C_m + V_{f,c}A\Delta C_m \qquad (21)$$

where $\dot{m}$, $D$, $\Delta C_m$, $U_{c,m}$ and $V_{f,c}$ represent rate of change of mass of the vaporizing droplet, coefficient of diffusion of the salt in the liquid, difference between the bulk and dynamic interfacial concentrations, the advection velocity due to the solutal gradient only, and velocity due to electro-magnetic-solutal advection, respectively. The internal circulation velocity due to the solutal gradient only is expressed similar to the thermal case as $U_{c,m} = \sigma_c \Delta C_m / \mu$, where $\sigma_c$ represents the rate of change of surface tension with respect to salt concentration.

Introducing the expression for the circulation velocity under field effect $V_{f,c}$ (similar to $V_f$ from eqn. (12)) in eqn. 21, the expression yields

$$\rho \dot{R} R = D\Delta C_m + \frac{\sigma_c}{\mu} R (\Delta C_m)^2 + \frac{\nu}{2R} \left\{ (HaI) \pm \sqrt{(HaI)^2 + 4E_{hd}} \right\} \Delta C_m \qquad (22)$$

$$\rho \dot{R} R = D\Delta C_m \left\{ 1 + Ma_s + \frac{Sc}{2} \left( HaI \pm \sqrt{(HaI)^2 + 4E_{hd}} \right) \right\} \qquad (23)$$

In eqn. (22- 23), $Sc$ represents the associated Schmidt number and $Ma_s$ represents the solutal Marangoni number, expressible as

$$Ma_s = \frac{\sigma_c R \Delta C_m}{D\mu} \qquad (24)$$

Rearranging further and applying the condition that $Ma_s \gg 1$ is essential for stable circulation, the expression can be obtained as

$$\rho \frac{\dot{R} R}{D\Delta C_m} = \left\{ Ma_s + \frac{Sc}{2} \left( HaI \pm \sqrt{(HaI)^2 + 4E_{hd}} \right) \right\} \qquad (25)$$

The term $\left[ Ma_s + \frac{Sc}{2} \left( HaI \pm \sqrt{(HaI)^2 + 4E_{hd}} \right) \right]$ acts as the effective electro-magnetic-solutal Marangoni number. The magnitude of this number represents the effective strength of the solutal advection within the vaporizing droplet as governed by the interacting electromagnetic fields.

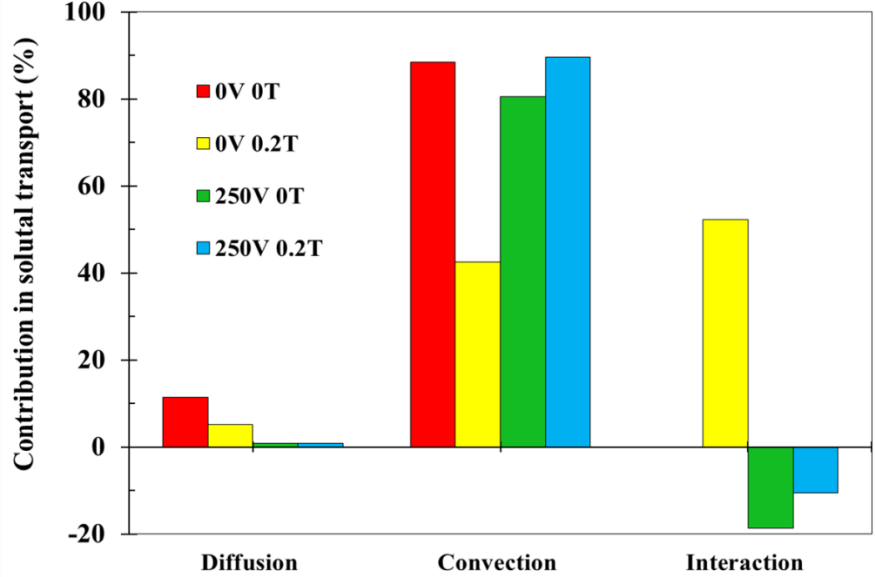

**FIG. 10:** The typical percentage contribution of the different mechanisms involved in the species transport during the vaporization of 0.2 M droplet in different electric and magnetic field constraints.

Similar to the thermal counterpart, the percentage contribution of the individual terms in the electro-magneto-solutal species balance scaling is illustrated in fig. 10. It comprises of the diffusion, convection, and the field interaction (individual fields and combined) terms. In the zero field case, the convection contribution ranges ~ 91%, and the diffusion contributes ~ 9%. In the case of only magnetic field, the magneto-solutal effect overshoots the convective component, and this conforms to previous reports [33]. In the case of only electric field, a reduction in the convective component is noted, with an opposing electro-advection component arising in the system [39]. However, despite the large magnitude of the magneto-advection (~ 55 %) compared to the electro-advection (~ –20%), the interaction component in the combined field system still poses an opposition to the internal advection. This furthers the proposition that in the interacting field system, the electrohydrodynamic advection is dominant over the ferrohydrodynamic counterpart, despite the latter being stronger independently. Further, it is also noted that the electro-magneto-thermal and solutal advection components are of nearly equal contribution to the overall internal hydrodynamics. Figure 11 illustrates a phase plot between the thermal ($Ma_T$) and solutal ($Ma_S$) Marangoni numbers. The stability regimes are represented by the iso-Lewis lines [54, 57, 58] and points lying farther to the right of the Le=0 represents enhanced stability of the internal advection due to solutal Marangoni effect compared to the thermal counterpart. It is noted that the points move farther to the right under various field constraints, indicating increased advection strength. However, this is in stark contrast to the experimental observations, and further probing is essential.

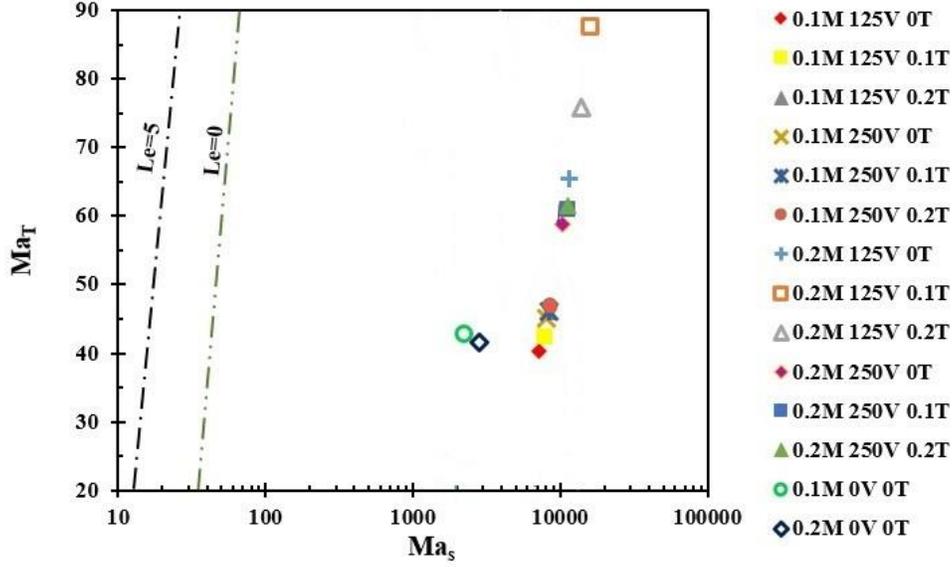

**FIG. 11:** Phase plot of the thermal ($Ma_T$) against the solutal Marangoni number ($Ma_s$) for vaporizing droplets under different electric and magnetic field constraints. The stability curves represent the different iso-$Le$ lines [54, 57].

Fig. 12 illustrates a similar phase map, albeit between the effective electro-magneto-thermal ($Ma_{T,m}$) and solutal ($Ma_{T,s}$) Marangoni numbers. This map incorporates the effects of the field interactions through the morphed Marangoni numbers. The iso-Ha and iso-$E_{hd}$ lines have been presented to understand the role played by the individual but independent fields. It is noted that while the increase in Ha leads to the points shifting towards improved solutal advection, the increase in the $E_{hd}$ leads to the points shifting towards reduced strength of solutal advection. It is also observed that under the different field interaction constraints, the points shift towards the regions of reduced advection, which is again in agreement with the experiments. The electro-magneto-thermal and electro-magneto-solutal advection velocities for different cases are predicted from the scaling analysis discussed previously. The scaling analysis also reveals that the net advection is due to competing effects of the two forms of advection. The average velocity of advection is thus expressed as

$$U_{c,m}\Big|_{effective} = \frac{1}{2}\left| U_{c,m}\Big|_{thermal} - U_{c,m}\Big|_{solutal} \right| \tag{26}$$

The values of the effective advection velocity as determined from the theoretical analyses have been illustrated in fig. 6 (e) against experimental velocimetry data, and accurate predictability has been noted. This thereby cements the hypothesis that the vaporization rates are morphed in accordance to the competitive interactions of the electro-magneto-thermal and solutal hydrodynamics.

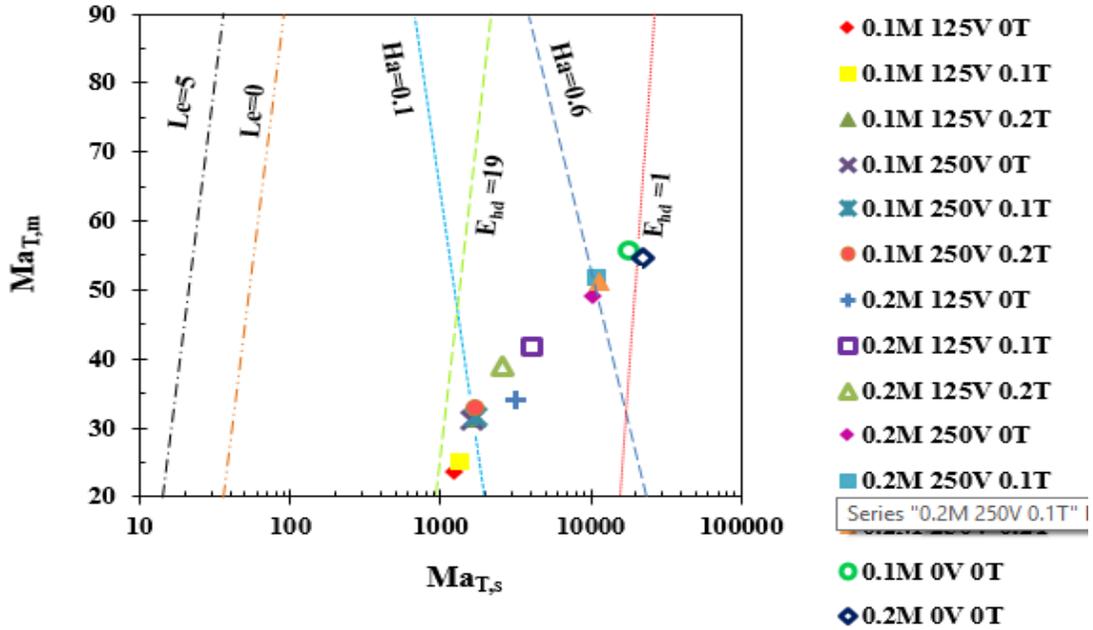

**Fig. 12:** Phase map of the electro-magneto-thermal ($Ma_{T,m}$) and electro-magneto-solutal ($Ma_{s,m}$) Marangoni number for different droplet configurations. The iso-$E_{hd}$ and iso-$Ha$ lines illustrate the stability regimes for the associated advection [39, 58].

## 4. Conclusions

The present article discusses the methodology and physics behind morphing and control of the vaporization rates of droplets employing induced electrohydrodynamics and ferrohydrodynamics, and their interactivities. Experiments with saline paramagnetic droplets with orthogonally placed electric and magnetic fields reveal that a wide gamut of vaporization rates can be realized by playing with the field strengths. The vapour diffusion model for evaporation is noted to be insufficient towards explaining the morphed vaporization rates. Likewise, the modulation of surface tension due to field is also not a justified mechanism to explain the observations. Velocimetry within the droplet reveals that the internal advection is morphed by the interacting electro and ferrohydrodynamics, which in turn controls the vaporization rates by interfacial shear. A theoretical analysis has been put forward to scale the energy and species transport mechanisms within the vaporizing droplet. The analysis reveals that the modulation of both the thermal and solutal interfacial and the internal advection by the fields is responsible for the noted advection kinetics. The role of the effective electro-magneto-thermal and electro-magneto-solutal Marangoni numbers has been discussed to this end. Phase stability analysis reveals that both the morphed thermal and solutal hydrodynamics is equally responsible towards the morphed advection patterns. The velocities predicted from the analysis are observed to conform accurately to the experimental velocities. Control and tuning of droplet vaporization rates can hold immense importance in several thermofluidic systems, across a wide range of length and time scales. The findings may have strong implications towards design, development and realization of such systems.